\documentclass[journal]{IEEEtran}
\usepackage{graphicx}
\usepackage{amsfonts, amsmath,amssymb}

\newcommand{\nc}{\newcommand}
\nc{\bsig}{\boldsymbol{\sigma}}
\nc{\bet}{\boldsymbol{\eta}}
\nc{\bfi}{\boldsymbol{\varphi}}
\nc{\bxi}{\boldsymbol{\xi}}

\begin{document}

\title{A Robust Scheme for PSS Detection and Integer Frequency Offset Recovery in LTE Systems}
\author{Michele Morelli, \textit{Senior Member, IEEE}, and Marco Moretti, 
\textit{Member, IEEE}
\thanks{\indent M. Morelli and M. Moretti are with the University of Pisa,
Dipartimento di Ingegneria dell'Informazione, Italy. Email:  \{michele.morelli, marco.moretti\}@iet.unipi.it.}
}
\maketitle

\begin{abstract}
Before establishing a communication link in a cellular network, the user terminal must activate a synchronization procedure called initial cell search in order to acquire specific information about the serving base station. To accomplish this task, the primary synchronization signal (PSS) and secondary synchronization signal (SSS) are periodically transmitted in the downlink of a Long Term Evolution (LTE) network.

Since SSS detection can be performed only after successful identification of the primary signal, in this work we present a robust scheme for joint PSS detection, sector index identification and integer frequency offset (IFO) recovery in an LTE system. The proposed algorithm relies on the maximum likelihood (ML) estimation criterion and exploits a suitable reduced-rank representation of the channel frequency response to take multipath distortions into account. We show that some PSS detection methods that were originally introduced through heuristic reasoning can be derived from our ML framework by selecting an appropriate model for the channel gains over the PSS subcarriers.

Numerical simulations indicate that the proposed scheme can be effectively applied in the presence of severe multipath propagation, where existing alternatives provide unsatisfactory performance.
\end{abstract}

\begin{keywords}
LTE, cell search, sector index identification, integer frequency offset
recovery.
\end{keywords}

\section{Introduction}

The Long Term Evolution (LTE) mobile communication standard has been
developed by the 3rd Generation Partnership Project (3GPP) in order to
enhance the performance of currently deployed 3G systems in terms of data
throughput, spectrum utilization and user mobility \cite{Dahlman2008}. This
technology\ supports various channel bandwidths ranging from 1.4 to 20 MHz
and promises peak data rates of 100 Mbit/s in the downlink (DL) and 50
Mbit/s in the uplink (UL) for the 20 MHz system bandwidth \cite{3GPP2009}.
Improved resilience against multipath distortion, high spectral efficiency
and ability to handle different data rates are achieved by using orthogonal
frequency-division multiple-access (OFDMA) in the DL, while single-carrier
frequency-division multiple-access (SC-FDMA) is adopted in the UL due to its
reduced peak-to-average power-ratio \cite{Myung2008}. There is also the
possibility of choosing between normal and extended cyclic prefix (CP), the
latter being considered for large delay spread environments.

LTE supports multi-cell communications, with cell information being conveyed
by an integer number called cell-ID.  Upon entering the network or during an handover operation, the user equipment
(UE) must recover the cell-ID of the serving base station (eNodeB) and has
also to acquire correct timing and frequency synchronization. This operation
is known as initial \textit{cell-search} \cite{Pin2000}, \cite{Wang2011} and
is accomplished by exploiting a dedicated synchronization channel (SCH)
periodically inserted in the DL radio frame \cite{3GPPPhy}. The SCH conveys
two signals called Primary Synchronization Sequence (PSS) and Secondary
Synchronization Sequence (SSS). The former is generated from a 63-length
frequency-domain Zafoff-Chu (ZC) sequence whose root index univocally
determines the sector identity. The latter is an interleaved concatenation
of two length-31 scrambled \textit{m}-sequences specifying the cell ID group.

Although in principle it is possible to consider the joint estimation of all
synchronization parameters and cell-ID, a more pragmatic approach relies on
the following three-stage procedure \cite{Yingming2007}:
\begin{enumerate}
\item Firstly, fractional frequency offset (FFO) and coarse symbol timing
recovery is accomplished using the redundancy introduced by the CP. This
method was originally proposed in \cite{Sandell1997} and its accuracy can be
improved by averaging the timing and frequency metrics over several OFDM
symbols;

\item In the second step, the UE detects the position of the PSS within the
received DL signal in order to acquire subframe timing information, and also
determines the sector index by identifying which primary
sequence has been transmitted out of three possible alternatives. The
integer frequency offset (IFO) can be retrieved at this stage by evaluating
the shift of the received PSS in the frequency domain;

\item The final step recovers the cell ID group and identifies the frame
boundary by using the received SSS. Once these operations have been
completed, the UE is able to read some basic configuration information
broadcast from the eNodeB, such as the system bandwidth, CP length and
duplexing mode.
\end{enumerate}

Since SSS detection can be accomplished only after successful identification
of the primary sequence, PSS detection represents a crucial task in the
overall cell search procedure. For this reason, it has attracted much
attention in the last few years and many solutions are currently available.
Some of them operate in the time-domain (TD), while others exploit the
frequency-domain (FD) samples provided by the receive discrete Fourier
transform (DFT) unit. Examples of TD schemes can be found in \cite{Setiawan2010}-\cite{Yu2013}, where the PSS is revealed by looking for the
peak of the cross-correlation between the received samples and the three
locally regenerated ZC sequences. However, since the SCH is transmitted onto
a set of 62 dedicated subcarriers with the other subcarriers being modulated
by data symbols, the PSS should be extracted from the DL signal before the
correlation stage. This requires a high-order filtering operation of the
received signal, which results into an increase of the hardware complexity
and also leads to some performance degradation since the contribution of
data subcarriers cannot be totally filtered out due to the spectral leakage
and the uncompensated CFO. TD methods with reduced complexity are suggested
in \cite{Zhang2012} and \cite{Gao2011}, where the tasks of PSS detection and
sector index recovery are decoupled by either exploiting the central
symmetric-property of the ZC sequences or by correlating the DL signal with
the sum of the three possible primary sequences.

As an alternative to the TD\ approach, in FD schemes the contribution of the
data-bearing subcarriers is eliminated by selecting the SCH at the receive
DFT output and correlating the resulting samples with replicas of the three
tentative frequency-domain ZC sequences \cite{Manolakis2009}. This method
works properly as long as the channel gain can be considered as
approximately constant over the SCH subband. Unfortunately, such an
assumption does not hold for transmissions over severe multipath channels or
in the presence of a non-negligible timing error, which appears as a
linearly increasing phase shift at the DFT output. In these circumstances,
the correlation properties of the received PSS will be irreparably destroyed
with an ensuing degradation of the system performance. To solve this
problem, differential correlation in the FD has been proposed in \cite{Tsai2012}, where IFO detection is also accomplished during the PSS match
process. A disadvantage of the differential approach is that the peaks of
the resulting metric are quite close for different IFO values, thereby
reducing its ability to recover the frequency error \cite{Su2013}.
Furthermore, compared to \cite{Manolakis2009}, the accuracy of the FD
differential correlator is greatly reduced whenever the channel frequency
response (CFR) over the SCH is approximately constant. For this reason, the
authors of \cite{Elsherif2013} propose to adaptively choose between
non-differential and differential detection on the basis of a previous
estimate of the maximum channel delay spread. A simpler solution is found in 
\cite{Wung2011} by resorting to the partial correlation concept, wherein the
received SCH is partitioned into several adjacent FD segments which are
subsequently correlated with the corresponding parts of the tentative ZC
sequences. A fundamental design parameter of this approach is represented by
the number of segments, which must be selected in accordance with the
coherence bandwidth of the transmission channel.

It is worth noting that all the aforementioned methods have been derived by
means of heuristic reasoning. In order to check whether their performance
can be substantially improved or not, it is of interest to make comparisons
with alternative approaches based on some optimality criterion. With this
goal in mind, in the present work we employ maximum likelihood (ML) methods
to study the problem of PSS detection, sector index identification and IFO
recovery, which represents the second step of the cell search process in an
LTE system. In contrast to previous investigations, our analysis, which originates from the work  presented in \cite{MoMo15}, explicitly
takes into account the multipath distortion introduced by the propagation
medium on the received signal. This is achieved by treating the CFR over the
SCH subcarriers as a nuisance vector \cite{MoMo2012}, which is jointly estimated along with
the synchronization parameters. As we shall see, this approach may lead to
an under-determined estimation problem wherein the number of quantities to
be recovered exceeds the number of available data. Reduced-rank
approximations of the CFR are proposed to cope with such a situation using
either the minimum mean square-error criterion (MMSE) or the classical
polynomial basis expansion. This approach results into a general framework
which includes both the conventional FD scheme \cite{Manolakis2009} and the
partial correlation concept \cite{Wung2011} through a suitable selection of
the CFR model. A key assumption of our analysis is that symbol timing and
FFO have been previously acquired using the classical CP-based method
presented in \cite{Sandell1997}. However, since multipath propagation may
significantly reduce the accuracy of the timing estimates provided by \cite{Sandell1997}, a residual timing error is included in the adopted system
model and numerical simulations are used to assess its impact on the overall
performance.

The rest of the paper is organized as follows. Next section illustrates the
signal model and formulates the estimation problem. Sect. III presents the
joint ML estimator of the unknown parameters using a generic expansion basis to represent the CFR. An MMSE reduced-rank
approximation of the CFR is derived in Sect. IV along with other possible
channel representations that lead to alternative known PSS detection
schemes. After discussing numerical simulations in Sect. V, we offer some
conclusions in Sect. VI.

\textit{Notation}: Matrices and vectors are denoted by boldface letters,
with $\mathbf{I}_{N}$ being the identity matrix of order $N$, $\mathbf{0}%
_{K} $ the $K$-dimensional null vector and $\mathbf{1}_{K}$ a $K$%
-dimensional vector with unit entries. The notation $\Vert \cdot\Vert
$ indicates the norm of the enclosed vector, while $\mathbf{B}^{-1}$ and tr$%
\{\mathbf{B}\}$ are the inverse and the trace of a matrix $\mathbf{B}$. We use E$%
\left\{ \cdot \right\} $, $(\cdot )^{\ast }$, $(\cdot )^{T}$ and $(\cdot
)^{H}$ for expectation, complex conjugation, transposition and Hermitian
transposition, respectively. The notation $\Re $e$\{\cdot \}$ stands for the
real part of a complex-valued quantity, while $|\cdot |$ represents the
corresponding modulus. Finally, we use $\tilde{\lambda}$ to indicate a trial
value of an unknown parameter $\lambda $.

\section{System model}

\subsection{LTE frame structure}

\begin{figure}[htbp]
\begin{center}
\includegraphics[width=8.5cm]{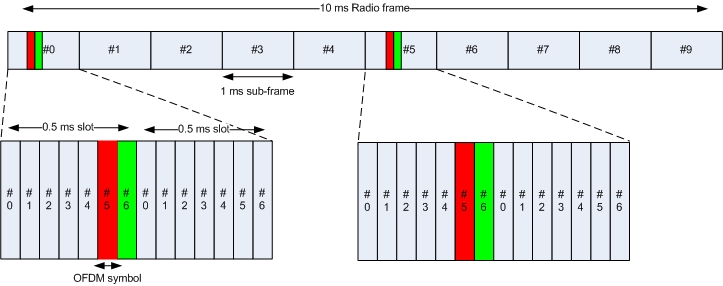}
\caption{Position of PSS (green) and SSS (red) in the LTE downlink frame.}
\label{fig1}
\end{center}
\end{figure}

We consider the DL of an LTE system operating in
frequency-division-duplexing (FDD) mode. As shown in Fig.~\ref{fig1}, data
transmission is organized in radio frame units of length 10 ms. Each frame
is divided into ten 1 ms subframes, which are further partitioned into two
slots of length 0.5 ms. Every slot contains 7 OFDMA symbols in case of
normal CP, while 6 OFDMA symbols are present when the extended CP is
employed to cope with a large delay spread. In the time-frequency grid,
system resources are divided into resource blocks (RBs), such that each RB
is composed of 12 contiguous subcarriers for a duration of one slot.
According to the LTE specifications, 504 different physical-layer cell IDs
are available and arranged into 168 distinct groups. Each group is identified
by the cell ID group $N_{ID}^{(1)}\in \{0,1,\ldots ,167\}$ and contains
three different sectors, which are specified by the sector ID $N_{ID}^{(2)}\in \{0,1,2\}$. In addition to acquiring the correct timing and
frequency synchronization, during the cell search procedure the UE must retrieve the integer-valued parameters $N_{ID}^{(1)}$ and $N_{ID}^{(2)}$,
from which the cell ID of the serving eNodeB is uniquely computed as $N_{ID}^{cell}=3N_{ID}^{(1)}+N_{ID}^{(2)}$. To accomplish these tasks, two
synchronization sequences called PSS and SSS are periodically transmitted on
a dedicated SCH to specify the sector ID and cell group, respectively. In
particular, the PSS is located in the last OFDM symbol of the first and 11th
slots of each radio frame, while the SSS is transmitted in the symbol
immediately preceding the PSS. As LTE allows for various system bandwidths,
both PSS and SSS are mapped onto a 73-subcarriers subband located
symmetrically around DC, which corresponds to the smallest allowable
spectrum occupancy. Such design choice enables the detection of the
synchronization signals without requiring explicit knowledge of the system
bandwidth.

\subsection{Primary synchronization sequence}

The PSS is chosen in a set of three different 63-length ZC sequences
characterized by excellent auto- and cross-correlation properties \cite{Popovic1992}. More precisely, during the PSS transmission the SCH
subcarriers with indices $n\in \{0,\pm 1,\pm 2,\ldots ,\pm 36\}$ are
modulated by the following polyphase complex exponential terms
\begin{equation}
a_{u}(n)=\left\{ 
\begin{array}{cc}
e^{-j\pi u(n^{2}+63n+110)/63} &\text{if } n\in \mathcal{I}\\ 
0 & \text{otherwise}
\end{array}
\right. 
\label{Eq1}
\end{equation}
where $u$ denotes the root index of the selected ZC sequence and $\mathcal{I}%
=\{n\in \mathbb{Z}:\left\vert n\right\vert \leq 31$ and $n\neq 0\}$. The
above equation indicates that, out of the 73 available subcarriers, only 62
are modulated by the PSS, while the remaining eleven (five placed at the SCH
boundaries and one placed at DC) are left unfilled. The root index belongs
to the set $\mathcal{J}_{u}=\left\{ 25,29,34\right\} $ and specifies the
sector ID $N_{ID}^{(2)}$ as shown in Table \ref{tab:ZCroots}.
\begin{table}[htbp] 
  \centering
  \caption{Relationship between $N_{ID}^{(2)}$ and ZC roots}\label{tab:ZCroots}
  \begin{tabular}{||c|c||}
        \hline
          $N_{ID}^{(2)}$ & Root index $u$\\
         \hline
        $0$&$25$\\
         \hline
        $1$&$29$\\
        \hline
       $2$&$34$\\
        \hline
        \end{tabular}
\medskip
\end{table}

\subsection{Overview of the cell search procedure}

Cell search is a three-stage procedure which is performed when the UE is
switched on or when it loses synchronization. Since LTE must provide
mobility up to 350 km/h, this process is periodically repeated to find a
candidate cell for the handover operation. In the first stage, the UE
retrieves the FFO and acquires coarse information about the OFDM symbol
timing. This operation is typically accomplished in the time domain using
the conventional CP-based delay correlation method presented in \cite{Sandell1997}. After FFO correction and CP removal, the resulting samples
are converted in the frequency domain using a DFT unit. The second stage
detects the position of the PSS within the received DL signal and recovers
the ZC root index $u$. These tasks can be accomplished either in the time or
frequency domain, and provide subframe timing information as well as the
sector ID. In the third stage, the SSS is exploited to get the cell ID group
and the frame boundary. Recalling that the SSS is located in the symbol
immediately preceding the PSS, the latter is normally used as a phase
reference to perform coherent detection of the SSS in the frequency domain.
As for the IFO, it can be estimated either in the second or third step by
evaluating the frequency domain shift of the received PSS or SSS at the DFT
output. However, since there are 168 different secondary sequences compared
to only three primary sequences, the complexity of the IFO search is greatly
reduced if it is performed in the second step in conjunction with PSS
detection.

\subsection{Signal model and problem formulation}

We now concentrate on the second stage of the cell search process and
provide the signal model for PSS detection and IFO\ recovery. The DL signal
propagates through a multipath channel with discrete-time impulse response $%
\mathbf{h}=[h(0),h(1),\ldots ,h(L-1)]^{T}$ of order $L$. At the UE receiver,
the incoming signal is down-converted to baseband and sampled with period $%
T_{s}$. Then, the CP delay correlation method \cite{Sandell1997} is applied
to retrieve the FFO and the OFDM symbol timing. After FFO compensation,
timing information is exploited to remove the CP and to divide the stream of
time-domain samples into segments $\mathbf{x}_{k}=[x_{k}(0),x_{k}(1),\ldots
,x_{k}(N-1)]$ $($with $k=1,2,\ldots )$ which are subsequently fed to the $N-$%
point DFT unit. We denote by
\begin{equation}
X_{k}(n)=\frac{1}{\sqrt{N}}\sum\limits_{\ell =0}^{N-1}x_{k}(\ell )e^{-j2\pi
\ell n/N}  \label{Eq2}
\end{equation}
the $n$th DFT output corresponding to the $k$th segment and assume that the
accuracy of the frequency and timing estimates is such that we can
reasonably neglect any inter-channel or inter-block interference present on $%
X_{k}(n)$. However, due to a possible residual timing error, the quantities $%
X_{k}(n)$ will be affected by a phase rotation that increases linearly with
the subcarrier index $n$. Since the PSS is transmitted twice per frame, the
observation window for PSS detection encompasses $N_{Q}$ adjacent OFDM
symbols composing a half frame, with $N_{Q}$ being either 60 or 70 for the
extended or normal CP transmission mode, respectively. Without loss of
generality, in this work we consider an LTE\ system with an extended CP and
let $N_{Q}=60$. Furthermore, we assume that the PSS is transmitted on the $q$%
th OFDM symbol within the observation window $k=1,2,\ldots ,N_{Q}$ and
denote by $\mathbf{X}_{k}=[X_{k}(-36),X_{k}(-35),\ldots ,X_{k}(36)]$ the
73-dimensional vector collecting the DFT output over the SCH channel. Then,
for $k\neq q$ we model $\mathbf{X}_{k}$ as a zero mean Gaussian vector with
covariance matrix $\mathbf{C}_{k}=\sigma _{k}^{2}\mathbf{I}_{73}$ ($\sigma
_{k}^{2}$ being an unknown parameter), while for $k=q$ we have
\begin{equation}
X_{q}(n)=H(n-\nu )a_{u}(n-\nu )e^{j2\pi (n-\nu )\theta /N}+w_{q}(n)\quad\left\vert n\right\vert \leq 36  
\label{Eq3}
\end{equation}
where $\nu $ represents the IFO, $\theta $ denotes the residual timing error
(normalized by the sampling period) and $H(n)$ is the CFR over the $n$th
subcarrier. Finally, $w_{q}(n)$ is the noise contribution, which is modeled
as a circularly-symmetric white Gaussian process with average power $\sigma
_{w}^{2}=$E$\{\left\vert w_{q}(n)\right\vert ^{2}\}$. Without loss of
generality, we consider $\theta $ as an integer-valued parameter. The reason
is that any residual fractional timing offset (FTO) can be incorporated into 
$h(\ell )$ by letting $h(\ell )=\left. h(t)\right\vert _{t=(\ell
+\varepsilon )T_{s}}$, where $\varepsilon $ denotes the FTO and $h(t)$ is
the continuous-time version of the channel impulse response (CIR).

Our goal is to exploit the set of observations $\mathbf{X}=[\mathbf{X}%
_{1}^{T}~\mathbf{X}_{2}^{T}~\ldots ~\mathbf{X}_{N_{Q}}^{T}]^{T}$ to find the
joint ML estimate of the unknown parameters $\{q,u,\nu \}$ specifying the
PSS position, the sector ID and the IFO. Unfortunately, this task is
complicated by the presence of the nuisance quantities $\boldsymbol{\{}%
\mathbf{H,}\theta ,\bsig^{2},\sigma _{w}^{2}\}$, where $\mathbf{H}%
=\{H(n);\left\vert n\right\vert \leq 31\}$ and $\bsig%
^{2}=\{\sigma _{k}^{2};1\leq k\leq N_{Q}$ and $k\neq q\}$.{\normalsize \ In
order to reduce the number of estimation parameters, we propose to merge }$%
\mathbf{H}$ and $\theta $ into an equivalent CFR $\mathbf{H}%
_{eq}=\{H_{eq}(n);\left\vert n\right\vert \leq 31\}$ with elements $%
H_{eq}(n)=H(n)e^{-j2\pi n\theta /N}$. Moreover, since {\normalsize channel
gains over blocks of contiguous subcarriers are highly correlated, we expect
that }$\mathbf{H}_{eq}$ {\normalsize can be accurately expanded over a }%
reduced-rank basis of $P<63$ vectors $\{\mathbf{b}_{1},\mathbf{b}_{2},\ldots
,\mathbf{b}_{P}\}$. This amounts to putting
\begin{equation}
\mathbf{H}_{eq}\simeq \mathbf{B\bxi }  \label{Eq4}
\end{equation}
where $\mathbf{B}$ is a matrix with columns $\mathbf{b}_{i}$ ($i=1,2,\ldots
,P$) and $\bxi=[\xi (1),\xi (2),\ldots ,\xi (P)]^{T}$ is a vector of expansion coefficients. A number of possible reduced-rank representations of 
$\mathbf{H}_{eq}$ will be presented later.

\section{Joint estimation of the unknown parameters}

We arrange the PSS position, sector ID and IFO into a vector $\bfi=[q,u,\nu ]$ and let $\bet=[\bxi,\bsig%
^{2},\sigma _{w}^{2}]$ be the set of nuisance parameters. The log-likelihood
function (LLF) for the unknown quantities $\{\bfi,\bet\}$ is obtained from (\ref{Eq3}) as
\begin{align}\label{Eq5}
\Lambda (\tilde{\bfi},\tilde{\bet})=&-\sum\limits_{\substack{ k=1  \\ k\neq \tilde{q}}}^{N_{Q}}\left[ 73\ln (\pi \tilde{\sigma}_{k}^{2})+\frac{1}{\tilde{\sigma}_{k}^{2}}\left\Vert \mathbf{X}_{k}\right\Vert ^{2}\right] \\
&-73\ln (\pi \tilde{\sigma}_{w}^{2})-\frac{1}{\tilde{\sigma}_{w}^{2}}\sum\limits_{n=-36}^{36}\left\vert X_{\tilde{q}}(n)-s(\tilde{\bfi},\tilde{\bxi};n)\right\vert ^{2} \nonumber
\end{align}
where we have defined $s(\tilde{\bfi},\tilde{\bxi};n)=%
\tilde{H}_{eq}(n-\tilde{\nu})a_{\tilde{u}}(n-\tilde{\nu})$ with $\tilde{%
\mathbf{H}}_{eq}=\mathbf{B}\tilde{\bxi}$. The joint ML estimate of
the unknown parameters is found by looking for the global maximum of $%
\Lambda (\tilde{\bfi},\tilde{\bet})$. Maximizing with
respect to $\tilde{\bsig}^{2}$ and $\tilde{\sigma}_{w}^{2}$
produces
\begin{equation}
\hat{\sigma}_{k}^{2}=\frac{1}{73}\left\Vert \mathbf{X}_{k}\right\Vert ^{2}
\label{Eq6}
\end{equation}
and 
\begin{equation}
\hat{\sigma}_{w}^{2}=\frac{1}{73}\sum\limits_{n=-36}^{36}\left\vert X_{\tilde{q}}(n)-s(\tilde{\bfi},\tilde{\bxi};n)\right\vert ^{2}.  
\label{Eq7}
\end{equation}
These results are substituted back into (\ref{Eq5}) in place of $\tilde{%
\sigma}_{k}^{2}$ and $\tilde{\sigma}_{w}^{2}$, yielding
\begin{align}
\Lambda (\tilde{\bfi},\tilde{\bxi})
=&73\left[-\sum\limits_{k=1}^{N_{Q}}\ln \left( \frac{\pi }{73}\left\Vert \mathbf{X}%
_{k}\right\Vert ^{2}\right) +\ln \left(\frac{\pi}{73} \left\Vert \mathbf{X}_{\tilde{q}}\right\Vert ^{2}\right)  \right. \\
& \left .- N_{Q}-\ln \left(\frac{\pi}{73} \sum\limits_{n=-36}^{36}\left\vert X_{\tilde{q}%
}(n)-s(\tilde{\bfi},\tilde{\bxi};n)\right\vert
^{2}\right)\right]. \notag
\end{align}
Skipping irrelevant multiplicative and additive terms independent of the
optimization variables and exploiting the monotonicity of the log function,
we may replace $\Lambda (\tilde{\bfi},\tilde{\bxi})$
by the equivalent objective function
\begin{equation}
\Lambda _{1}(\tilde{\bfi},\tilde{\bxi})=-\frac{%
\sum\limits_{n=-36}^{36}\left\vert X_{\tilde{q}}(n)-\tilde{H}_{eq}(n-\tilde{%
\nu})a_{\tilde{u}}(n-\tilde{\nu})\right\vert ^{2}}{\left\Vert \mathbf{X}_{_{%
\tilde{q}}}\right\Vert ^{2}}.  \label{Eq9}
\end{equation}
To proceed further, we make a change of the indexing variable $n-\tilde{\nu}%
\rightarrow m$ and recall that the PSS values $a_{u}(m)$ are non-zero only
for $m\in \mathcal{I}$ as specified in (\ref{Eq1}). Hence, assuming that $%
\left\vert \tilde{\nu}\right\vert \leq 5$, after dropping an immaterial term
in (\ref{Eq9}) we obtain
\begin{equation}
\Lambda _{2}(\tilde{\bfi},\tilde{\bxi})=\frac{2\Re 
\text{e}\left\{ \sum\limits_{m=-31}^{31}Z_{\tilde{q}}(\tilde{u},\tilde{\nu}%
;m)\tilde{H}_{eq}^{\ast }(m)\right\} -\left\Vert \tilde{\mathbf{H}}_{eq}\right\Vert ^{2}}{\left\Vert \mathbf{X}_{_{\tilde{q}%
}}\right\Vert ^{2}}.  \label{Eq10}
\end{equation}
where we have defined $Z_{\tilde{q}}(\tilde{u},\tilde{\nu};m)=X_{\tilde{q}%
}(m+\tilde{\nu})a_{\tilde{u}}^{\ast }(m)$. Denoting by $\mathbf{Z}_{\tilde{q}%
}(\tilde{u},\tilde{\nu})$ the 63-dimensional vector with entries $\{Z_{%
\tilde{q}}(\tilde{u},\tilde{\nu};m);\left\vert m\right\vert \leq 31\}$, we
may express $\Lambda _{2}(\tilde{\bfi},\tilde{\bxi})$
in matrix notation as
\begin{equation}
\Lambda _{2}(\tilde{\bfi},\tilde{\bxi})=\frac{2\Re 
\text{e}\left\{ \tilde{\bxi}^{H}\mathbf{B}^{H}\mathbf{Z}_{\tilde{q}%
}(\tilde{u},\tilde{\nu})\right\} -\tilde{\bxi}^{H}\mathbf{B}^{H}%
\mathbf{B}\tilde{\bxi}}{\left\Vert \mathbf{X}_{_{\tilde{q}%
}}\right\Vert ^{2}}.  \label{Eq11}
\end{equation}
The maximum of $\Lambda _{2}(\tilde{\bfi},\tilde{\bxi}%
) $ with respect to $\tilde{\bxi}$ is achieved at
\begin{equation}
\hat{\bxi}=\left( \mathbf{B}^{H}\mathbf{B}\right) ^{-1}\mathbf{B}%
^{H}\mathbf{Z}_{\tilde{q}}(\tilde{u},\tilde{\nu})  \label{Eq12}
\end{equation}
and plugging this result back into (\ref{Eq11}) yields the concentrated
likelihood function
\begin{equation}
\Lambda _{3}(\tilde{\bfi})=\frac{\mathbf{Z}_{\tilde{q}}^{H}(%
\tilde{u},\tilde{\nu})\mathbf{GZ}_{\tilde{q}}(\tilde{u},\tilde{\nu})}{%
\left\Vert \mathbf{X}_{_{\tilde{q}}}\right\Vert ^{2}}  \label{Eq13}
\end{equation}
where $\mathbf{G}=\mathbf{B}\left( \mathbf{B}^{H}\mathbf{B}\right) ^{-1}%
\mathbf{B}^{H}$. The joint ML estimate of the unknown parameters is
eventually obtained as
\begin{equation}
\hat{\bfi}=\arg \underset{\tilde{\bfi}}{\max }%
\{\Lambda _{3}(\tilde{\bfi})\}  \label{Eq14}
\end{equation}
which requires a search over the multi-dimensional domain spanned by $\tilde{%
\bfi}$. However, since $\tilde{\bfi}=[\tilde{q},%
\tilde{u},\tilde{\nu}]$ has integer-valued entries, there is only a finite
number of hypothesized values of $\tilde{\bfi}$. At this stage,
the problem arises as how to select a suitable basis $\{\mathbf{b}_{1},%
\mathbf{b}_{2},\ldots ,\mathbf{b}_{P}\}$ for the reduced-rank representation
of $\mathbf{H}_{eq}$ which can exhibit low-sensitivity to the timing error $%
\theta $. Possible solutions are presented in the next Section.

\section{Reduced-rank representation of the channel frequency response}

\subsection{Problem formulation}

In OFDM systems, the CFR is typically expressed in terms of the
discrete-time CIR as 
\begin{equation}
H(n)=\sum_{\ell =0}^{L-1}h(\ell )e^{-j2\pi n\ell /N}.  \label{Eq15}
\end{equation}
Recalling that $H_{eq}(n)=H(n)e^{-j2\pi n\theta /N}$, from (\ref{Eq15}) it
follows that the equivalent CFR takes the form
\begin{equation}
H_{eq}(n)=\sum_{\ell =\theta }^{L+\theta -1}h(\ell -\theta )e^{-j2\pi n\ell
/N}.  \label{Eq16}
\end{equation}
Since coarse timing recovery schemes for OFDM systems adopt a back-off
design wherein the timing estimates are pre-advanced to avoid
inter-block-interference at the DFT output \cite{Minn2003}, in the sequel it
is assumed that there is no negative timing error. This amounts to letting $%
\theta $ vary in the set $\{0,1,\ldots ,\theta _{\max }\}$, where $\theta
_{\max }$ is selected on the basis of the accuracy of the timing estimator.
In such a case, the equivalent CFR can be written in matrix notation as
\begin{equation}
\mathbf{H}_{eq}=\mathbf{Fh}_{eq}  \label{Eq17}
\end{equation}
where $\mathbf{h}_{eq}=[\mathbf{0}_{\theta }^{T}$ $\ \mathbf{h}^{T}$ \ $%
\mathbf{0}_{\theta _{\max }-\theta }^{T}]^{T}$ is the equivalent CIR vector
of order $L_{eq}=L+\theta _{\max }$ and $\mathbf{F}$ is a $63\times L_{eq}$
matrix with entries
\begin{equation}
\lbrack \mathbf{F}]_{n,\ell }=e^{-j2\pi n\ell /N}\quad\left\vert
n\right\vert \leq 31,\text{ \ }0\leq \ell \leq L_{eq}-1.  \label{Eq18}
\end{equation}
Since in a practical LTE scenario the normalized timing error may be as large as 40 
\cite{Su2013}, the channel order $L_{eq}$ is expected to be close or even to 
exceed the value 63. In this case, we cannot interpret (\ref{Eq17}) as a
reduced-rank representation of $\mathbf{H}_{eq}$ and the goal is to find a
basis $\{\mathbf{b}_{1},\mathbf{b}_{2},\ldots ,\mathbf{b}_{P}\}$ of order $%
P\ll 63$ such that the orthogonal projection of $\mathbf{H}_{eq}$ on the
selected basis, say $\mathbf{H}_{P}=\mathbf{GH}_{eq}$, is a good
approximation of $\mathbf{H}_{eq}$. The accuracy of such an approximation
can be measured in terms of the mean-square-error (MSE) between $\mathbf{H}%
_{eq}$ and $\mathbf{H}_{P}$, which is defined as E$\{\left\Vert \mathbf{H}%
_{eq}-\mathbf{H}_{P}\right\Vert ^{2}\}$. Recalling that $\mathbf{G}=\mathbf{B%
}\left( \mathbf{B}^{H}\mathbf{B}\right) ^{-1}\mathbf{B}^{H}$, from (\ref%
{Eq17}) the MSE is found to be
\begin{equation}
\text{MSE}(\mathbf{B)}=\text{E}\{\mathbf{h}_{eq}^{H}\mathbf{F}^{H}\mathbf{G}%
^{\bot }\mathbf{Fh}_{eq}\}  \label{Eq19}
\end{equation}
where we have taken into account that $\mathbf{G}^{\bot }=\mathbf{I}_{63}-%
\mathbf{G}$ is an idempotent matrix and we have explicitly indicated the
dependence of the MSE on the expansion matrix $\mathbf{B}$. After standard
manipulations, we can rewrite (\ref{Eq19}) as
\begin{equation}
\text{MSE}(\mathbf{B)}=\text{tr}\{\mathbf{G}^{\bot }\mathbf{FC}_{eq}\mathbf{F%
}^{H}\}  \label{Eq20}
\end{equation}
with $\mathbf{C}_{eq}=$E$\{\mathbf{h}_{eq}\mathbf{h}_{eq}^{H}\}$ being the
covariance matrix of $\mathbf{h}_{eq}$.

\subsection{MMSE reduced-rank representation}

For a fixed value of $P$, the optimum expansion basis is the one that
minimizes MSE$(\mathbf{B)}$. Neglecting an irrelevant additive term
independent of $\mathbf{B}$, it is seen that the minimum of MSE$(\mathbf{B)}$
is achieved by maximizing the metric
\begin{equation}
\gamma (\mathbf{B)}=\text{tr}\{\mathbf{B}\left( \mathbf{B}^{H}\mathbf{B}%
\right) ^{-1}\mathbf{B}^{H}\mathbf{FC}_{eq}\mathbf{F}^{H}\}  \label{Eq21}
\end{equation}
with respect to $\mathbf{B}$. To solve this problem, it is convenient to
consider the compact singular value decomposition (SVD) of $\mathbf{B}$,
which is given by
\begin{equation}
\mathbf{B}=\mathbf{U\Sigma V}^{H}  \label{Eq22}
\end{equation}
where $\mathbf{\Sigma \in \mathbb{R}}^{P\times P}$ is the diagonal matrix containing the $P$ non-zero singular
values of $\mathbf{B}$ sorted in decreasing order, with the columns of $\mathbf{U\in\mathbb{C}}^{63\times P}$ and $\mathbf{V\in\mathbb{C}}^{P\times P}$ being the corresponding left and right eigenvectors, respectively. Substituting (\ref{Eq22}) into (\ref{Eq21}) and observing that $\mathbf{U}^{H}\mathbf{U}=\mathbf{V}^{H}\mathbf{V=I}_{P}$, after some algebraic computations we can rewrite $\gamma (\mathbf{B)}$ as
\begin{equation}
\gamma (\mathbf{B)}=\text{tr}\{\mathbf{U}^{H}\mathbf{FC}_{eq}\mathbf{F}^{H}%
\mathbf{U}\}  \label{Eq23}
\end{equation}
from which we see that the objective function only depends on $\mathbf{U}$.
Maximizing the right-hand-side of (\ref{Eq23}) with respect to $\mathbf{U}$
is a well-known optimization problem, whose solution is obtained by
selecting as columns of $\mathbf{U}$ the $P$ eigenvectors of $\mathbf{FC}%
_{eq}\mathbf{F}^{H}$ associated to the $P$ largest eigenvalues. In the
sequel, we denote by $\mathbf{U}_{MMSE}$ the matrix $\mathbf{U}$
provided by such a design criterion. Since $\mathbf{\Sigma }$ and $\mathbf{V}
$ can be arbitrarily chosen without affecting the value of the MSE, for
simplicity we let $\mathbf{\Sigma }=\mathbf{V=I}_{P}$ \ and obtain $\mathbf{%
B=U}_{MMSE\text{ }}$.

Unfortunately, computing $\mathbf{U}_{MMSE\text{ }}$ requires knowledge of $%
\mathbf{C}_{eq}$, which depends on the channel power delay profile and the
timing error. Since these quantities are generally unknown at the receiver,
the exact MMSE solution cannot be pursued in practice. A possible way out is
found by allowing the system to operate in a mismatched mode wherein $%
\mathbf{C}_{eq}$ is replaced by $\mathbf{I}_{L_{eq}}$. This leads to an
approximated MMSE (AMMSE) solution, say $\mathbf{B=U}_{AMMSE\text{ }}$, in
which the columns of $\mathbf{B}$ are the $P$ normalized eigenvectors of $\mathbf{FF}%
^{H}$ associated to the $P$ largest eigenvalues. In such a case, the
concentrated likelihood function in (\ref{Eq13}) becomes 
\begin{equation}
\Lambda _{3}(\tilde{\bfi})=\frac{\left\Vert \mathbf{U}_{AMMSE%
\text{ }}^{H}\mathbf{Z}_{\tilde{q}}(\tilde{u},\tilde{\nu})\right\Vert ^{2}}{%
\left\Vert \mathbf{X}_{_{\tilde{q}}}\right\Vert ^{2}}.  \label{Eq24}
\end{equation}

\subsection{Polynomial-based reduced-rank (PRR) representation}

An alternative reduced-rank representation of $\mathbf{H}_{eq}$ can be
obtained by approximating the CFR with a $(P-1)-$order polynomial function as
\begin{equation}
H_{eq}(n)\simeq \sum_{p=1}^{P}\xi (p)n^{p-1}\quad\left\vert
n\right\vert \leq 31.  \label{Eq25}
\end{equation}
This amounts to putting $\mathbf{B}=\mathbf{B}_{PRR}$, where $\mathbf{B}_{PRR}$ is a $63\times P$ matrix with entries
\begin{equation}
\lbrack \mathbf{B}_{PRR}]_{n,p}=n^{p-1}\quad\left\vert
n\right\vert \leq 31,\text{ \ }1\leq p\leq P.  \label{Eq26}
\end{equation}
\subsection{Piecewise-constant reduced-rank (PCRR) representation}

The piecewise-constant reduced-rank representation (PCRR) of the CFR is
obtained by arranging the 63 SCH subcarriers into $P$ adjacent subbands and
assuming that $H_{eq}(n)$ is approximately constant on each subband. Hence,
denoting by $K_{p}$ the number of subcarriers contained in the $p$th subband
(with $p=1,2,\ldots ,P$), we have
\begin{equation}
H_{eq}(n)\simeq \xi (p)\quad J_{p-1}-31\leq n\leq J_{p}-32
\label{Eq27}
\end{equation}
where
\begin{equation}
J_{p}=\sum_{m=1}^{p}K_{m}\quad 1\leq p\leq P  \label{Eq28}
\end{equation}
and $J_{0}=0$. If $P$ is an integer divider of 63, all subbands have the
same number of subcarriers $K_{p}=63/P$. Otherwise, we let $63=MP+R$, where $%
M=$int$\{63/P\}$ and $R\in \{1,2,\ldots ,P-1\}$ are the quotient and
remainder of the integer division $63/P$, respectively. Then, the size of the $P$ subbands
are designed such that $K_{p}=M+1$ for $1\leq p\leq R$ and $K_{p}=M$ for $%
R+1\leq p\leq P$ . In matrix notation, $\mathbf{H}_{eq}$ can be written as
in (\ref{Eq4}) after letting $\mathbf{B=B}_{PCRR\text{ }}$ ,where $\mathbf{B}%
_{PCRR\text{ }}$ is the $63\times P$ following matrix 
\begin{equation}
\mathbf{B}_{PCRR\text{ }}=\left[ 
\begin{array}{cccc}
\mathbf{1}_{K_{1}} & \mathbf{0}_{K_{1}} & \cdots  & \mathbf{0}_{K_{1}} \\ 
\mathbf{0}_{K_{2}} & \mathbf{1}_{K_{2}} & \cdots  & \mathbf{0}_{K_{2}} \\ 
\vdots  & \vdots  & \ddots  & \vdots  \\ 
\mathbf{0}_{K_{P}} & \mathbf{0}_{K_{P}} & \cdots  & \mathbf{1}_{K_{P}}
\end{array}
\right]   \label{Eq.29}
\end{equation}
In such a case, the concentrated likelihood function in (\ref{Eq13}) becomes 
\begin{equation}
\Lambda _{3}(\tilde{\bfi})=\frac{1}{\left\Vert \mathbf{X}_{\tilde{q}}\right\Vert ^{2}}\sum_{p=1}^{P}\frac{1}{K_{p}}\left\vert
\sum_{n=J_{p-1}-31}^{J_{p}-32}Z_{\tilde{q}}(\tilde{u},\tilde{\nu};n)\right\vert ^{2}  
\label{Eq30}
\end{equation}
which is reminiscent of the partial correlation method presented in \cite%
{Wung2011}. Interestingly, letting $P=1$ in (\ref{Eq30}) yields
\begin{equation}
\Lambda _{3}(\tilde{\bfi})=\frac{1}{63\left\Vert \mathbf{X}_{\tilde{q}}\right\Vert ^{2}}\left\vert \sum_{n=-31}^{31}Z_{\tilde{q}}(\tilde{u},\tilde{\nu};n)\right\vert ^{2}  \label{Eq31}
\end{equation}
which coincides with the conventional frequency-domain correlation-based
(CFDC) metric originally proposed in \cite{Manolakis2009} for the joint
estimation of $u$ and $\nu $.

\subsection{Complexity issues}

The processing load of the estimator (\ref{Eq14}) depends on which
reduced-rank representation is adopted for the CFR. For both the AMMSE and
PRR solutions, the metric $\Lambda _{3}(\tilde{\bfi})$ can be
efficiently computed as
\begin{equation}
\Lambda _{3}(\tilde{\bfi})=\frac{\left\Vert \mathbf{C}^{H}%
\mathbf{B}^{H}\mathbf{Z}_{\tilde{q}}(\tilde{u},\tilde{\nu})\right\Vert ^{2}}{%
\left\Vert \mathbf{X}_{_{\tilde{q}}}\right\Vert ^{2}}  \label{Eq32}
\end{equation}
where $\mathbf{CC}^{H}=\left( \mathbf{B}^{H}\mathbf{B}\right) ^{-1}$ is the
Choleski decomposition of $\left( \mathbf{B}^{H}\mathbf{B}\right) ^{-1}$,
while for the PCRR it is convenient to evaluate $\Lambda _{3}(\tilde{\bfi})$ as indicated in (\ref{Eq30}). In any case, for each new
received OFDM symbol, the receiver must compute the quantities $\left\Vert 
\mathbf{X}_{_{\tilde{q}}}\right\Vert ^{2}$ and $\mathbf{Z}_{\tilde{q}}(%
\tilde{u},\tilde{\nu})$. Assuming that the entries of $\mathbf{X}_{_{\tilde{q%
}}}$ are available, 291 floating-point-operations (flops) are required to
get $\left\Vert \mathbf{X}_{_{\tilde{q}}}\right\Vert ^{2}$, while 372 flops
are needed to evaluate $\mathbf{Z}_{\tilde{q}}(\tilde{u},\tilde{\nu})$ for
each couple $(\tilde{u},\tilde{\nu})$. When using the AMMSE solution, the
numerator of $\Lambda _{3}(\tilde{\bfi})$ in (\ref{Eq32}) is
computed with $498P$ flops for each $(\tilde{u},\tilde{\nu})$. This figure
reduces to $250P$ flops with the PRR approximation as in this case the
matrix $\mathbf{C}^{H}\mathbf{B}^{H}$ is real-valued. As for the PCRR,
evaluating $\Lambda _{3}(\tilde{\bfi})$ in (\ref{Eq30})
starting from $\left\Vert \mathbf{X}_{_{\tilde{q}}}\right\Vert ^{2}$ and $%
\mathbf{Z}_{\tilde{q}}(\tilde{u},\tilde{\nu})$ needs $123+3P$ flops for each
couple $(\tilde{u},\tilde{\nu})$, which reduces to $126$ flops when
considering the CFDC. The overall complexity of the investigated schemes is
summarized in Table \ref{tab:complexity} for each received OFDM symbol. In writing these
figures we have borne in mind that $\tilde{u}$ varies in the set $\left\{
25,29,34\right\} $ and we have denoted by $N_{\nu }$ the number of different
values of $\tilde{\nu}$.
\begin{table}[htbp] 
  \centering
  \caption{Complexity of the investigated schemes in terms of number of flops for each OFDM symbol.}\label{tab:complexity}
  \begin{tabular}{||c|l||}
        \hline
          Algorithm & Required flops\\
          \hline
        AMMSE&$291+1116N_{\nu}+1494N_{\nu}P$\\
         \hline
        PRR&$291+1116N_{\nu}+750N_{\nu}P$\\   
             \hline
       PCRR&$291+1485N_{\nu}+9N_{\nu}P$\\
        \hline
         CFDC&$291+1494N_{\nu}$\\
        \hline
        \end{tabular}
\medskip
\end{table}
\section{Simulation results}

Computer simulations have been run to assess the performance of the
presented PSS detection and IFO recovery schemes using different
reduced-rank representations of the CFR. The LTE simulation set-up is chosen
according to the 3GPP specifications \cite{3GPPPhy} and is summarized as
follows.

\subsection{Simulation model}

We consider a 20 MHz LTE communication system with 15 kHz subcarrier
spacing. At the receiver side, the baseband signal is sampled with frequency 
$f_{s}=30.72$ MHz and converted in the frequency-domain through a 2048-point
DFT unit. To demonstrate the capability of the investigated scheme in a
challenging scenario, we adopt the Extended Typical Urban (ETU) channel
model characterized by 9 channel taps with maximum excess delay $\tau _{\max
}=5$ $\mu $s. The path gains are modeled as statistically independent random
variables with zero-mean and Gaussian distribution (Rayleigh fading). A
raised-cosine functions with roll-off 0.22 and time-duration of 6 sampling
periods is employed for the pulse shaping, which corresponds to an overall
CIR of order $L=$int$\{f_{s}\tau _{\max }+6\}=160$. Assuming a maximum
normalized timing error $\theta _{\max }=40$ \cite{Su2013}, the length of
the equivalent CIR vector $\mathbf{h}_{eq}$ is found to be $L+\theta _{\max
}=200$. However, since parameter $P$ (and the system complexity) is expected
to increase with the CIR\ duration, in the design of the AMMSE we use $L_{eq}=120$ , which amounts to reducing the size of matrix $\mathbf{F}$ from 
$63\times 200$ to $63\times 120$. This choice is motivated by the fact that
in the ETU channel model there is only one multipath component with a path
delay greater than $2.3$ $\mu $s which, moreover, collects less than 3\% of
the average channel power. The value $L_{eq}=120$ is also compliant with
other LTE channel models, such as the Extended Vehicular A (EVA) and
Extended Pedestrian A (EPA), which are characterized by a maximum excess
delay of $2.51$ $\mu $s and $0.41$ $\mu $s, respectively.

Without any loss of generality, we adopt the extended CP transmission mode
wherein 6 OFDM symbols are present in each slot. Since the PSS is
transmitted every 10 slots, in our simulations we let $N_{Q}=60$. The range
of CFO values is related to the oscillator instability, while it is only
marginally affected by the UE mobility. Hence, assuming that the stability
of commercial oscillators for mobile applications is in the order of $\pm 10$
parts-per-million (ppm) at both the transmit and receive ends, the maximum
CFO value is approximately $2.66$ subcarrier spacing at the carrier
frequency of 2 GHz. Accordingly, the search range for the IFO is $\mathcal{J}%
_{\nu }=\left\{ 0,\pm 1,\pm 2,\pm 3\right\} $, which amounts to putting $%
N_{\nu }=7$. Recalling that the PSS is chosen from a set of three possible
ZC sequences with root index $u\in \left\{ 25,29,34\right\} $, the overall
search space for the triplet $\tilde{\bfi}=[\tilde{q},\tilde{u},%
\tilde{\nu}]$ has cardinality $N_{\bfi}=3N_{q}N_{\nu }=1260$.

The accuracy of the PSS detection and IFO recovery schemes is measured in
terms of the error rate incurred in the estimation of each parameter of
interest $q,u$ and $\nu $. As a global performance indicator, we also
consider the overall probability of failure $P_{f}=\Pr \{\hat{\bfi}\neq \bfi\}$.

\subsection{Performance evaluation}

\begin{figure}[htbp]
\begin{center}
\includegraphics[width = 8.5cm]{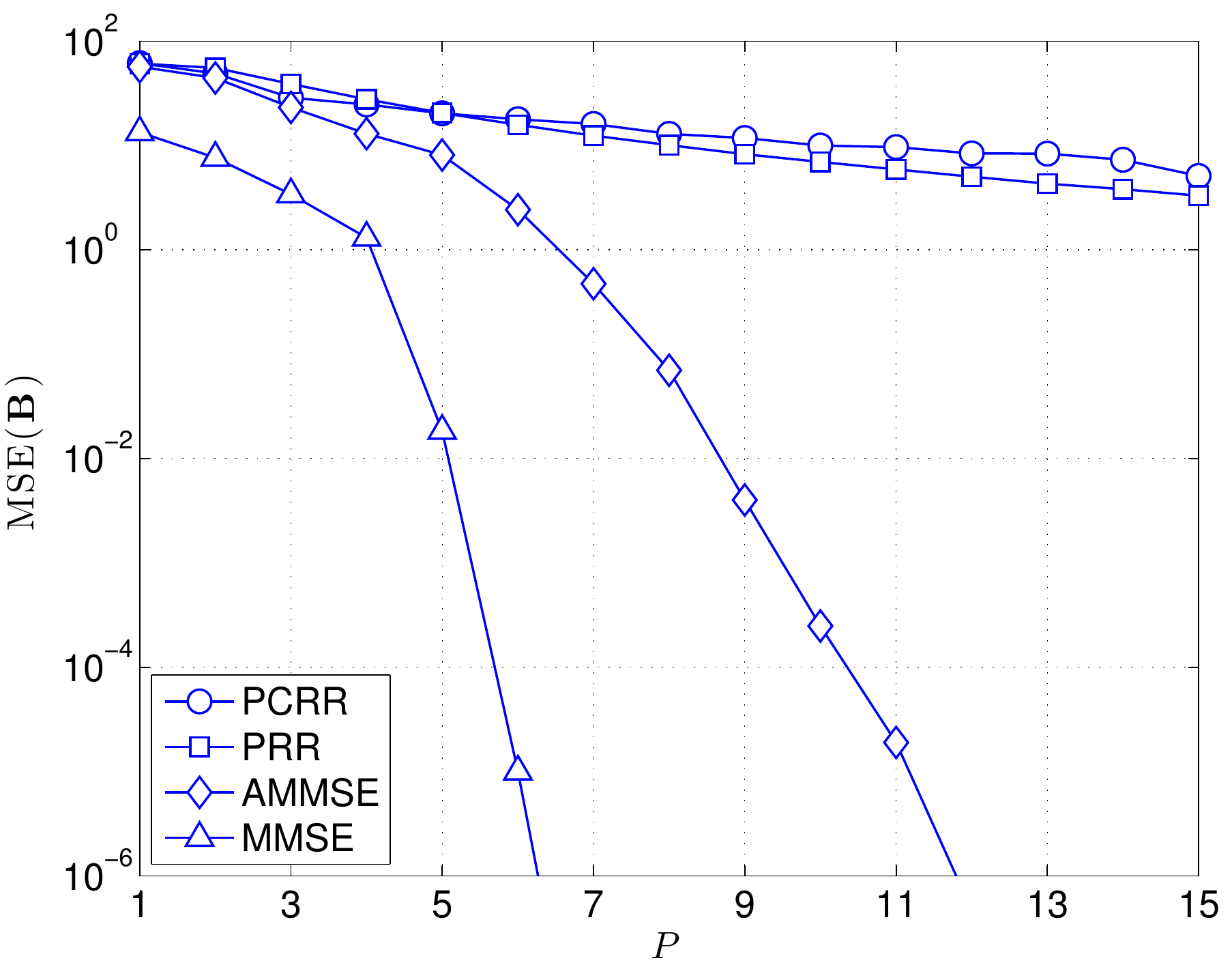}
\caption{MSE$(\mathbf{B)}$ vs. $P$ for different reduced-rank CFR representations.}
\label{fig2}
\end{center}
\end{figure}

Fig.~\ref{fig2} illustrates the quantity MSE$(\mathbf{B)}$ reported in (\ref{Eq20})
as a function of $P$ for the investigated reduced-rank CFR representations.
As expected, the best results are obtained with $\mathbf{B=U}_{MMSE\text{ }}$
since this choice minimizes MSE$(\mathbf{B)}$ for a fixed value of $P$.
Compared to the true MMSE solution, the AMMSE is characterized by a higher
MSE, even though it largely outperforms both the PRR and PCRR schemes.
Although the accuracy of the considered reduced-rank representations
steadily improves with $P$, the probability of failure $P_{f}$ cannot
exhibit a similar behaviour. The reason is that $\mathbf{B}$ has dimension $%
63\times P$ and, accordingly, it becomes a square matrix when $P=63$. In the
latter case, $\mathbf{G}=\mathbf{B}\left( \mathbf{B}^{H}\mathbf{B}\right)
^{-1}\mathbf{B}^{H}$ reduces to $\mathbf{I}_{63}$ and $\mathbf{G}^{\bot }=%
\mathbf{I}_{63}-\mathbf{G}$ is therefore the null matrix. This situation has
two consequences. On one hand, from (\ref{Eq20}) we see that MSE$(\mathbf{B)}%
=0$, thereby justifying the fact that the MSE decreases as $P$ approaches
the value 63. On the other hand, substituting $\mathbf{G}=\mathbf{I}_{63}$
into (\ref{Eq13}) leads to 
\begin{equation}
\Lambda _{3}(\tilde{\bfi})=\frac{\left\Vert \mathbf{Z}_{\tilde{q}}(\tilde{u},\tilde{\nu})\right\Vert ^{2}}{\left\Vert \mathbf{X}_{_{\tilde{q}}}\right\Vert ^{2}}  \label{Eq33}
\end{equation}
or, equivalently,
\begin{equation}
\Lambda _{3}(\tilde{\bfi})=\frac{\sum\limits_{m=-31}^{31}\left
\vert X_{\tilde{q}}(m+\tilde{\nu})\right\vert ^{2}}{\left\Vert \mathbf{X}_{
\tilde{q}}\right\Vert ^{2}}.  \label{Eq34}
\end{equation}
This equation indicates that the metric $\Lambda _{3}(\tilde{\bfi})$ becomes independent of $\tilde{u}$ when $P=63$, thereby preventing any possibility to recover the sector index. From the above discussion it
follows that parameter $P$ must be judiciously designed so as to meet two
conflicting requirements. On one hand, it must be large enough to produce a
sufficiently accurate reduced-rank representation of the CFR. On the other
hand, it must be adequately small since otherwise the estimation problem
contains too many unknown quantities which cannot be estimated reliably.
\par
\begin{figure}[h]
\begin{center}
\includegraphics[width=8.5cm]{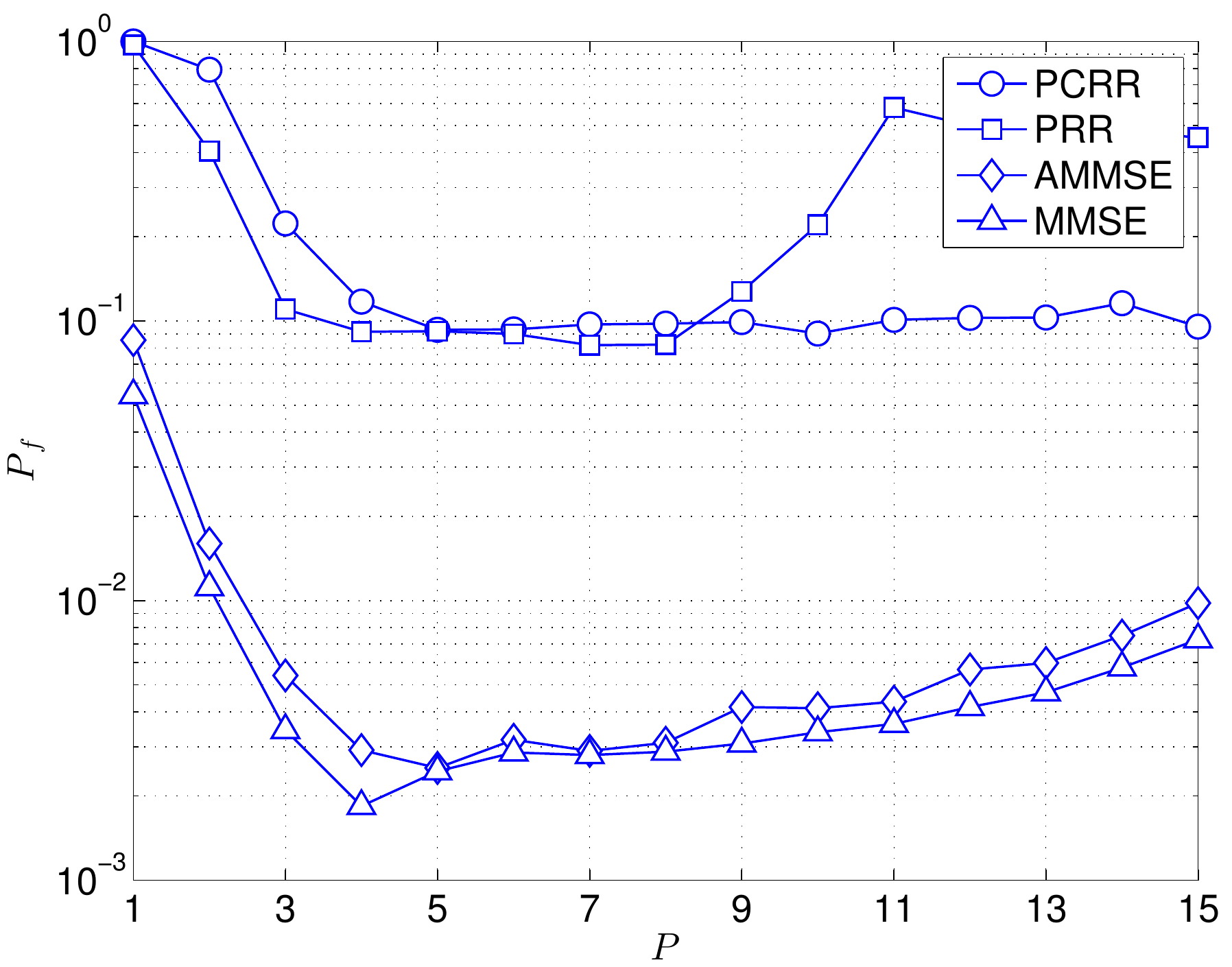}
\caption{Probability of synchronization failure vs. $P$ for $\text{SNR} =8$ dB and $\theta=40$.} 
\label{fig3}
\end{center}
\end{figure}

This intuition is corroborated by the results of Fig.~\ref{fig3}, where $P_{f}$ is
shown as a function of $P$. Here, the signal-to-noise ratio (SNR) is fixed
at 8 dB and the timing error is $\theta =\theta _{\max }=40$. As is seen,
for each curve there is an optimum value of $P$ which minimizes $P_{f}$.
Since the minimum is approximately attained at $P=5$ by all the considered
reduced-rank representations, such a value is used in the subsequent
simulations except for the CFDC scheme, which is obtained from the PCRR by
letting $P=1$.
\par
\begin{figure}[h]
\begin{center}
\includegraphics[width=8.5cm]{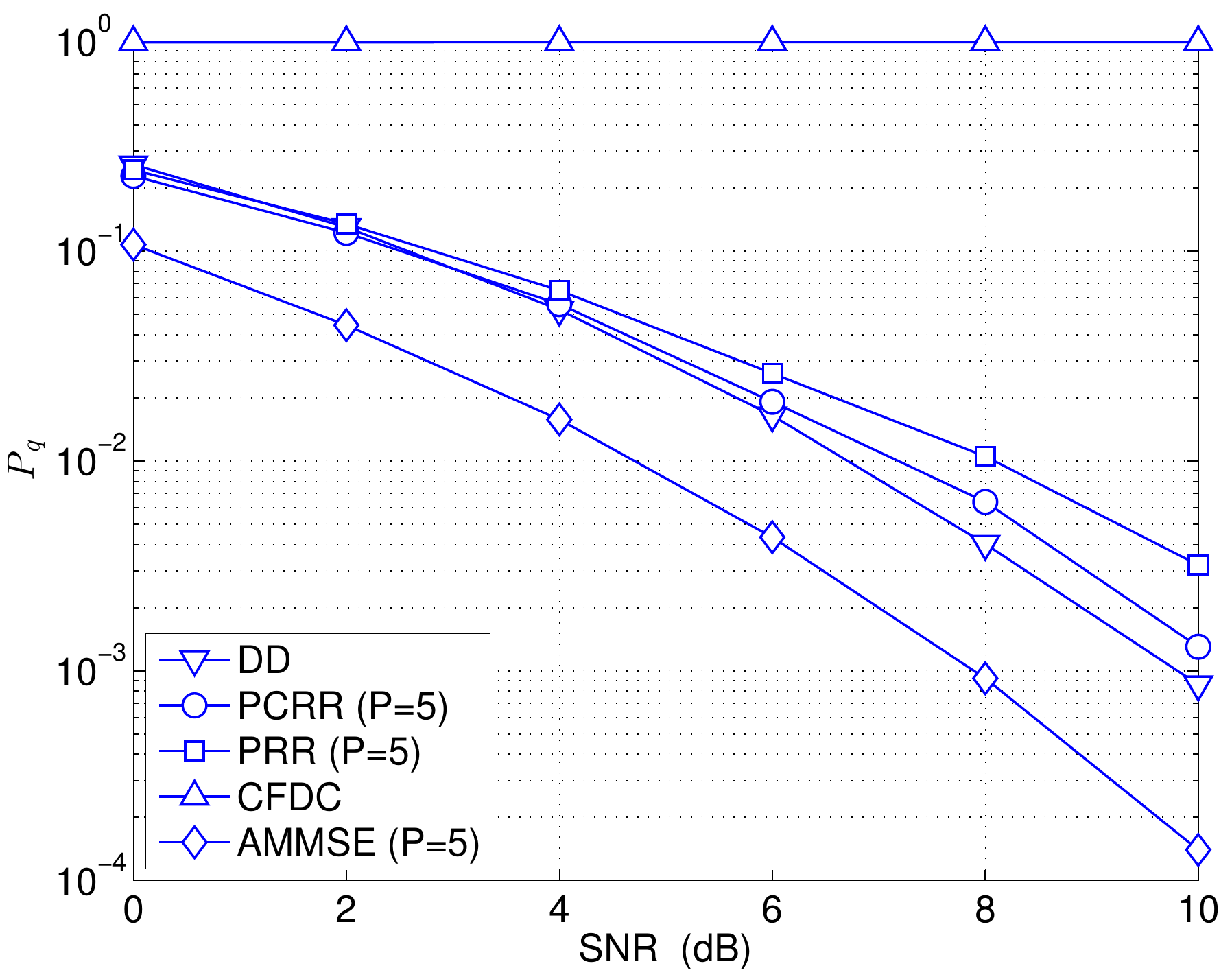}
\caption{Error rate incurred in the detection of the PSS vs. SNR for $\theta = 40$.}
\label{fig4}
\end{center}
\end{figure}

Fig.~\ref{fig4} illustrates the error rate incurred in detecting the presence of the
PSS, say $P_{q}=\Pr \{\hat{q}\neq q\}$, vs. the SNR for $\theta =40$. In
addition to the investigated schemes, we also consider the PSS differential
detector (DD) proposed in \cite{Manolakis2009}, which provides an estimate
of the unknown parameters $[q,u,\nu ]$ by looking for the maximum of the
following metric
\begin{equation}
\Lambda _{DD}(\tilde{\bfi})=\frac{\Re \text{e}\left\{
\sum\limits_{n=-30}^{31}Z_{\tilde{q}}(\tilde{u},\tilde{\nu};n)Z_{\tilde{q}}^{\ast }(\tilde{u},\tilde{\nu};n-1)\right\} }{\left\Vert \mathbf{X}_{\tilde{q}}\right\Vert ^{2}}  
\label{Eq35}
\end{equation}
in which the quantities $Z_{\tilde{q}}(\tilde{u},\tilde{\nu};n)$ and $Z_{\tilde{q}}^{\ast }(\tilde{u},\tilde{\nu};n-1)$ obtained from two adjacent
frequency bins are multiplied in order to mitigate the impact of the timing
error and channel selectivity on the system performance. As expected, AMMSE
outperforms all the other methods and achieves a gain of approximately 2 dB
over the DD, which further increases to 2.5 dB and 3 dB with respect to PCRR
and PRR, respectively. Due to its high sensitivity to timing errors and
channel distortions, the CFDC does not work properly in such adverse
scenario, and exhibits an error rate which is close to one at all SNR values.
\par
\begin{figure}[h]
\begin{center}
\includegraphics[width=8.5cm]{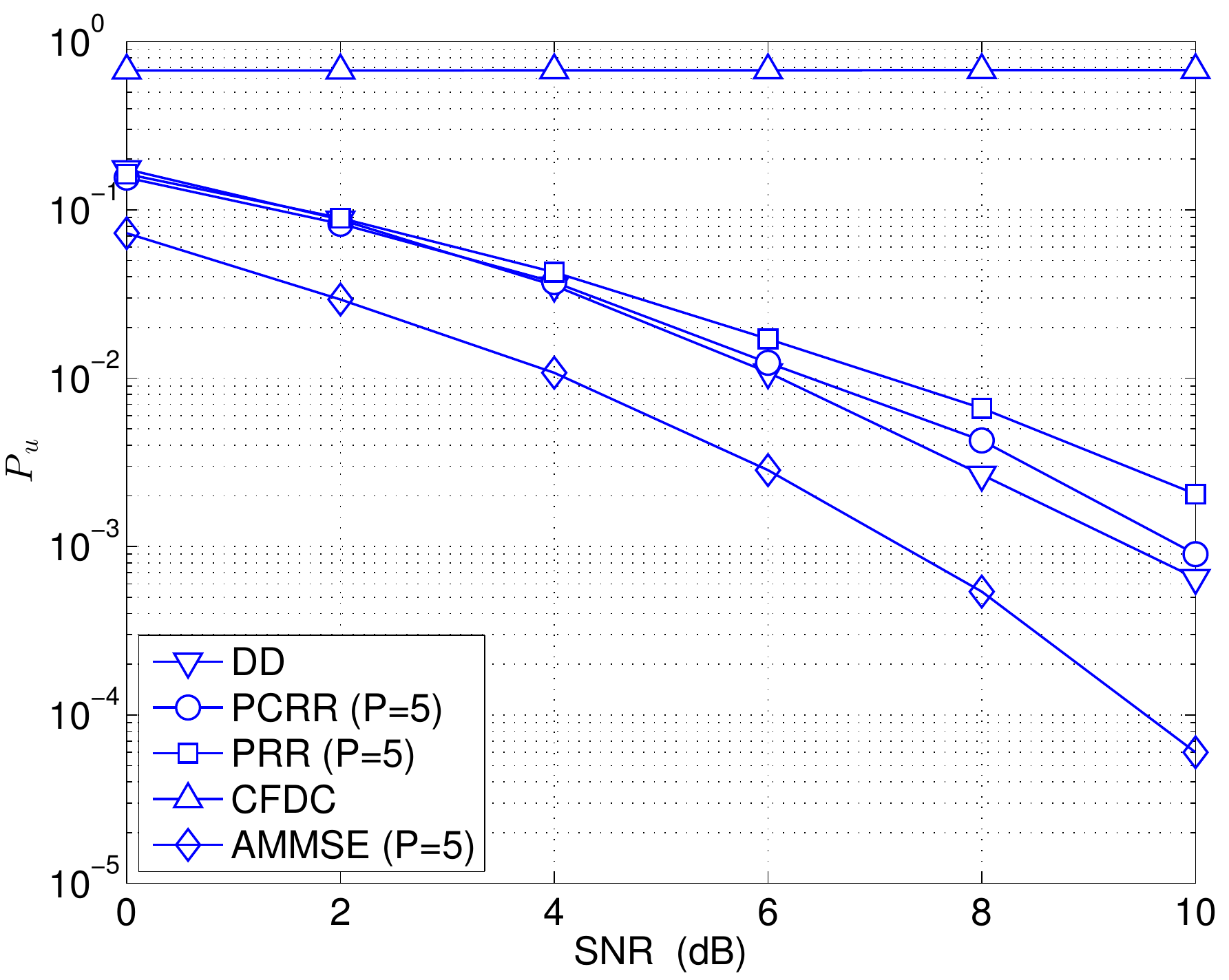}
\caption{Error rate incurred in the detection of the sector index $u$ vs. SNR for $\theta = 40$.}
\label{fig5}
\end{center}
\end{figure}
\begin{figure}[h]
\begin{center}
\includegraphics[width=8.5cm]{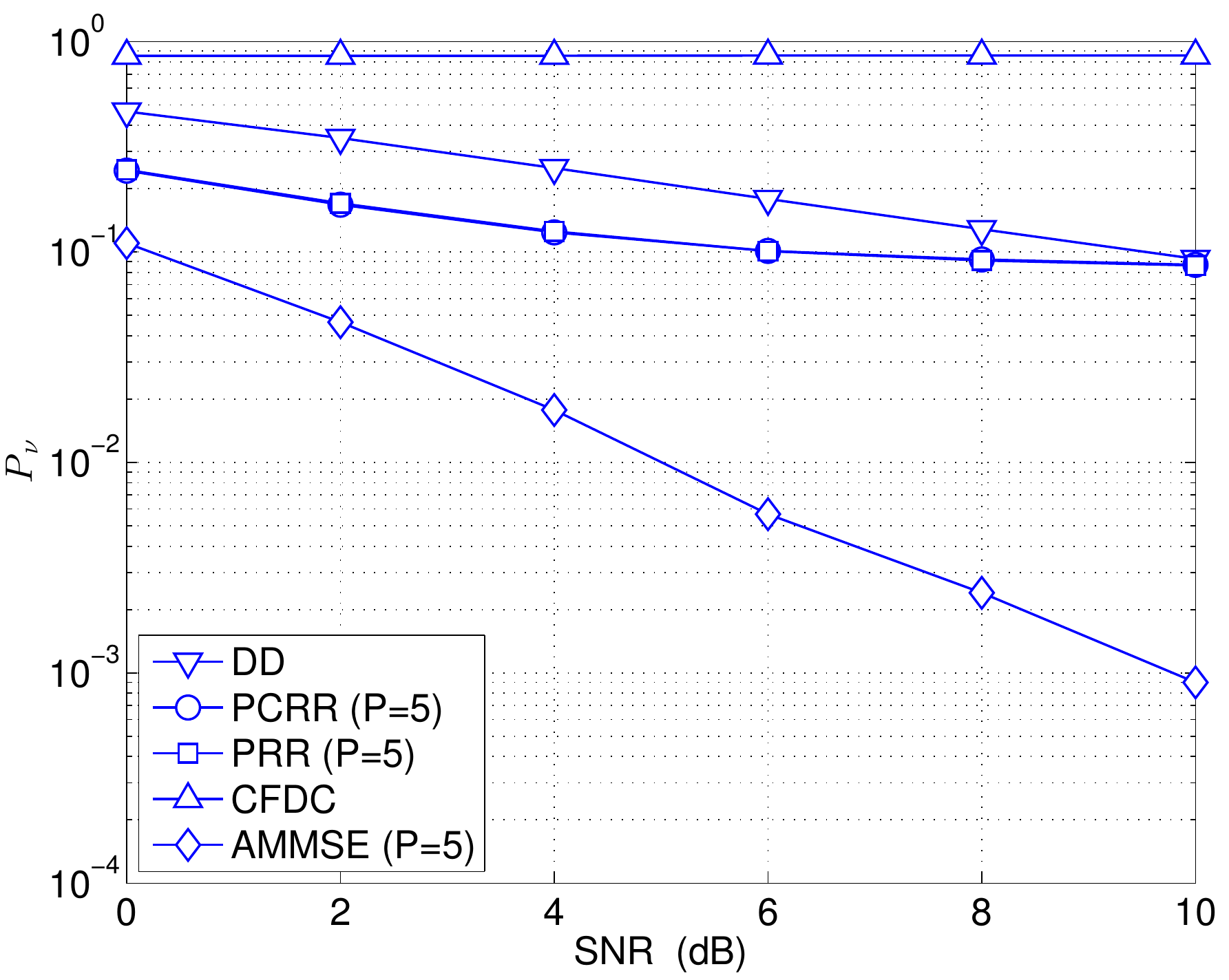}
\caption{Error rate incurred in the detection of the IFO $\nu$ vs. SNR for $\theta = 40$.}
\label{fig6}
\end{center}
\end{figure}

Figs.~\ref{fig5} and \ref{fig6} show the error rate incurred in the estimation of the sector
index and IFO, respectively, say $P_{u}=\Pr \{\hat{u}\neq u\}$ and $P_{\nu}=\Pr \{\hat{\nu}\neq \nu \}$. The operating scenario is the same as that in Fig.~\ref{fig4} and also
the trend of the curves is similar, with AMMSE outperforming the other
schemes. It is worth noting that the use of AMMSE is more advantageous for
the estimation of the IFO than for the other two parameters $q$ and $u$.
\par
\begin{figure}[htbp]
\begin{center}
\includegraphics[width=8.5cm]{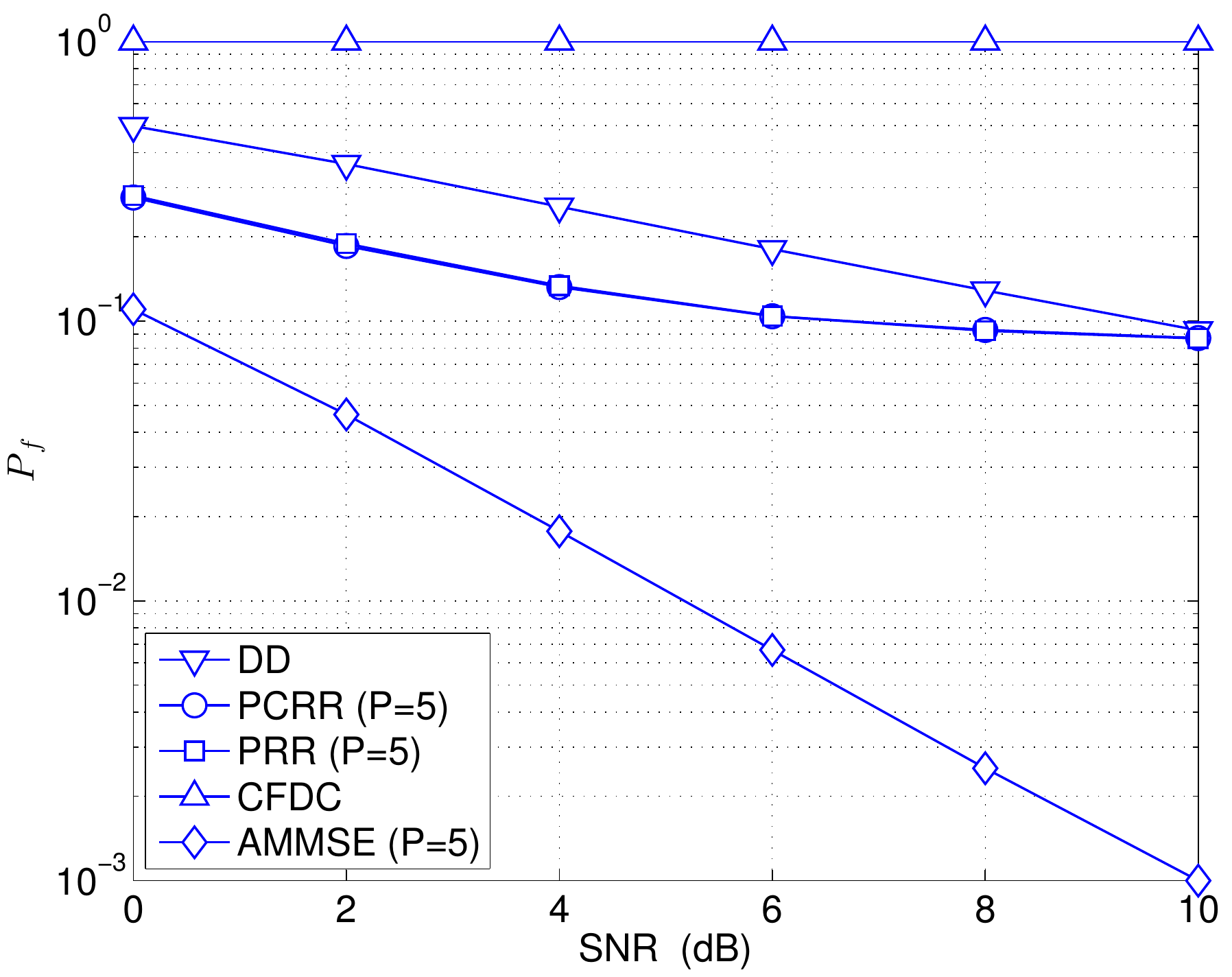}
\caption{Probability of synchronization failure vs. SNR for $\theta = 40$.}
\label{fig7}
\end{center}
\end{figure}

The overall system performance is summarized in Fig.~\ref{fig7}, where $P_{f}$ is
shown as a function of the SNR with the timing error being fixed to $\theta
=40$. These measurements validate the results of Figs.~\ref{fig4}-\ref{fig6} and clearly
indicate the superiority of AMMSE over the other methods. In particular, at
SNR=10 dB we see that the probability of failure\ of AMMSE is $10^{-3}$,
while it increases to $10^{-1}$ with PCRR, PRR and DD.
\par
\begin{figure}[htbp]
\begin{center}
\includegraphics[width=8.5cm]{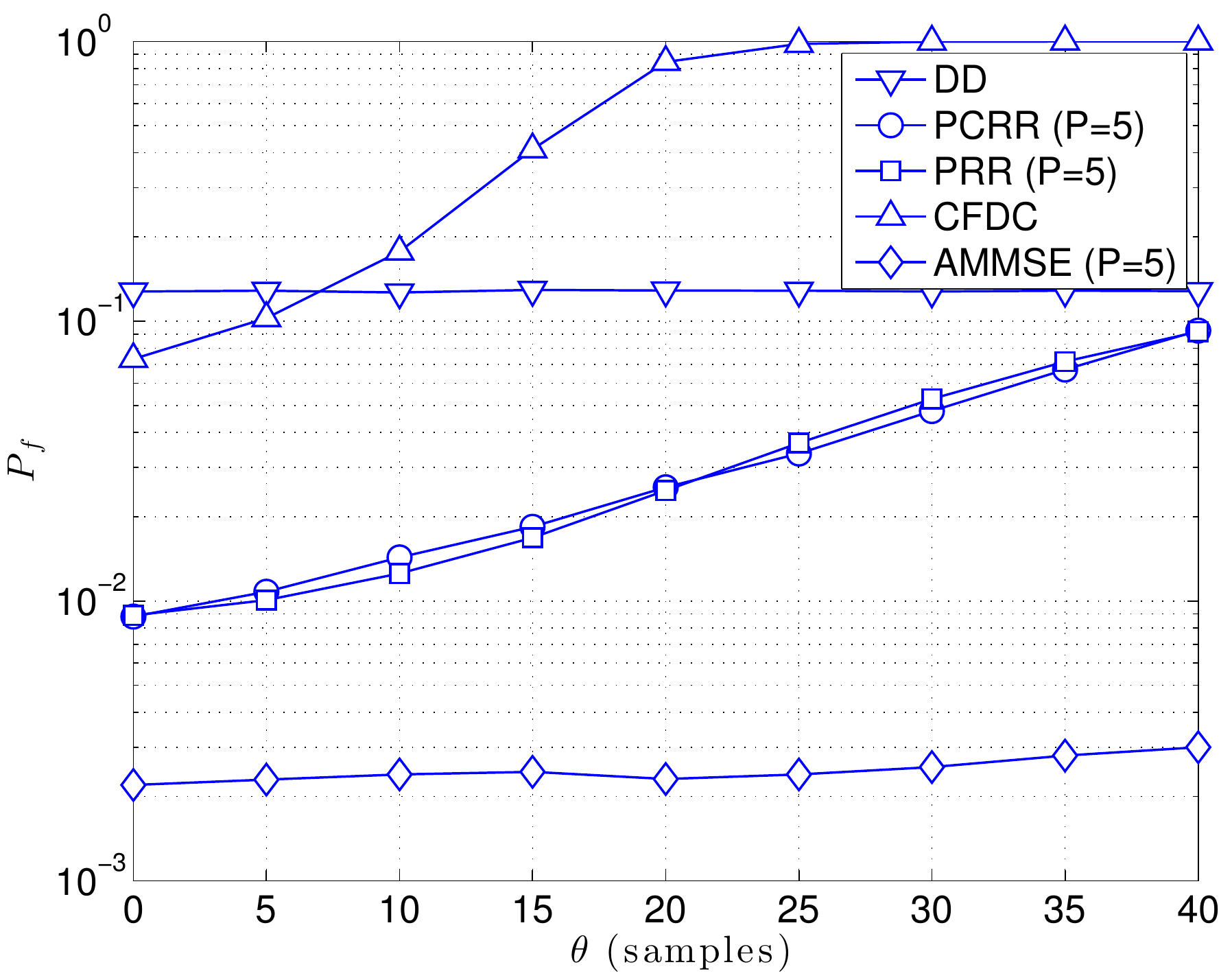}
\caption{Probability of synchronization failure vs. $\theta$  for $\text{SNR} =8$ dB.}
\label{fig8}
\end{center}
\end{figure}

Fig.~\ref{fig8} illustrates $P_{f}$ as a function of $\theta $ when the SNR is fixed
to 8 dB. These results are useful to assess the sensitivity of the
considered schemes to residual timing errors and demonstrate the remarkable
robustness of both DD and AMMSE to such an impairment. In contrast, the
probability of failure of PCRR, PRR and CFDC rapidly deteriorates as $\theta 
$ increases. This is particularly evident for the CFDC, which cannot be used
when the timing error exceeds a few sampling periods.

\subsection{Complexity comparison}

We conclude our study by comparing the investigated schemes in terms of
their computational complexity. From the results reported in Table \ref{tab:complexity} it
turns out that, in the considered scenario with $N_{\nu }=7$ and $P=5$, the
number of required kiloflops (kflops) for each received OFDM\ symbol is
approximately $60$ for the AMMSE method, $34$ for PRR and $11$ for both PCRR
and CFDC. As for the DD scheme, computing the numerator of $\Lambda _{DD}(%
\tilde{\bfi})$ in (\ref{Eq35}) starting from $\mathbf{Z}_{%
\tilde{q}}(\tilde{u},\tilde{\nu})$ needs 245 flops for each couple $(\tilde{u%
},\tilde{\nu})$ which, after including the computational burden required to
get the quantities $\left\Vert \mathbf{X}_{_{\tilde{q}}}\right\Vert ^{2}$
and $\mathbf{Z}_{\tilde{q}}(\tilde{u},\tilde{\nu})$, leads to an overall
processing requirement of $13$ kflops. These figures indicate that the
improved accuracy of the AMMSE is achieved at the price of a higher system
complexity. In particular, the number of flops is increased by a factor of
two with respect to PCRR, and by a factor 5.5 with respect to the other
methods. However, considering the fact that in the investigated simulation
set-up the AMMSE is the only scheme that can ensure acceptable error rate
performance, the penalty in terms of computational load seems tolerable.

\section{Conclusions}

We have presented an ML approach for joint PSS detection, sector index
identification and IFO recovery in the downlink of an LTE system. The
proposed scheme (AMMSE) operates in the frequency-domain and is based on a
novel MMSE reduced-rank representation of the channel frequency response
over the SCH subcarriers, which mitigates the impact of multipath
distortions and residual timing errors on the system performance. Compared
to existing schemes available in the literature, our method exhibits
improved accuracy at the price of a higher computational complexity. The penalty in
terms of required flops is justified by the fact that AMMSE can be used in a
harsh environment characterized by prolonged delay spreads and
non-negligible timing errors, where other competing schemes provide
unacceptable performance. Furthermore, it represents a promising candidate
for future heterogeneous networks in which the coexistence of femto-, pico-
and macro-cells will require fast and successful detection of neighboring
cells for efficient interference management and fast handover operations.

\end{document}